# Spectral Bandwidth Reduction of Thomson Scattered Light by Pulse Chirping

Isaac Ghebregziabher, Bradley Shadwick, and Donald Umstadter

*Department of Physics and Astronomy*
*University of Nebraska-Lincoln*

Compiled: April 2, 2012

Based on single particle tracking in the framework of classical Thomson scattering with incoherent superposition, we developed a fully relativistic, three dimensional numerical code that calculates and quantifies the characteristics of emitted radiation when a relativistic electron beam collides head-on with a focused counter-propagating intense laser field. The developed code has been benchmarked against analytical expressions, based on the plane wave approximation to the laser field, derived in *(1)*. For sufficiently long duration laser pulses, we find that the scattered radiation spectrum is broadened due to interferences arising from the pulsed nature of the laser. We show that by appropriately chirping the scattering laser pulse, the spectral broadening could be minimized.

## Introduction

Intense, tunable, ultra-short, collimated, polarized, and quasi-mono-energetic radiation in the x-ray and gamma-ray region of the electromagnetic spectrum has potential applications in broad disciplines that extends to, but is not limited to, natural sciences and health sciences. Such sources, which may be referred to as laser synchrotron sources (LSS), may be realized with the inverse Compton scattering, or Thomson scattering, process *(2-4)*. In inverse-Compton scattering, an intense laser field with frequency $\omega_0$ is scattered by a counter-propagating high energy ($E_e$) electron beam. In the limit where $E_e \gg 4\gamma^2 \hbar \omega_0$, the scattering process may be understood in terms of a classical Doppler shift and, for non-relativistic scattering laser intensities, the backscattered photon energy, $E_{sc}$, scales quadratically with the electron energy, i.e., $E_{sc} = 4\gamma^2 \hbar \omega_0$, where $\omega_0$ is the angular frequency of the scattering laser pulse, $\hbar$ is the reduced Plank's constant, and $\gamma$ the relativistic Lorentz factor. Unlike Bremsstrahlung radiation sources *(5)*, which are typically un-polarized, with broad energy spectrum, LSS sources have a high degree of polarization *(6)*, which is determined by the polarization of the scattering laser pulse, with relatively narrow band spectrum.

The generation of bright picosecond duration x-rays and gamma-rays through inverse-Compton scattering has already been demonstrated with head-on collision of intense laser pulses synchronized to picosecond-duration, high-energy electron beams generated with conventional radio frequency (RF) based accelerators *(7, 8)*. Ultra-short x-rays, with moderate brightness, have also been demonstrated from the inverse-Compton scattering of femto-second lasers in a $90^0$ scattering geometry with a synchronized, RF-accelerated, high-energy electron beam *(3)*. In addition to the experimental demonstration of x-ray and gamma-ray sources, the backscattered radiation was used as a diagnostic tool for the electron beam *(9, 10)* to generate polarized positrons *(11)* from dense targets, and to demonstrate nuclear fluorescence (proving the usefulness of the source for discerning isotope specific elements) *(12, 13)*.

The theory of Thomson backscattering of an infinitely long electromagnetic field by a relativistic electron beam has been well documented *(14-16)*. Moreover, extensive theoretical simulations have also been performed for scattering from a pulsed plane wave electromagnetic field. We extend these previous works by including the realistic 6-dimensional nature of the electron beam as well as the three-dimensional nature of the focused electromagnetic pulse with curved wave fronts. This treatment includes spectral broadening due to wave front curvature and finite temporal duration of the scattering laser and broadening associated with the transverse and longitudinal emittances of the electron beam. Such a detailed calculation of Thomson scattering is necessary to provide a framework for experimental events and to guide the design of Thomson-scattered x-ray sources.

In this paper, we discuss a numerical code that calculates the scattered radiation during the interaction of an intense laser pulse with an electron beam. In addition to benchmarking the code against previously reported results, we use it to demonstrate a technique to reduce the spectral bandwidth of Thomson scattered light by means of chirping the incident scattering laser pulse. The paper is organized as follows: Section 1 discusses the core ingredients of the developed code; in Section 2, a comparison of the numerically calculated and analytically obtained radiation energy is made; Section 3 discusses the effect of finite temporal width of the laser on the scattered spectrum as well as a method to overcome broadening due to the pulsed nature of the scattering laser. Summarized results are presented in Section 4.

## Modeling

The three-dimensional and fully relativistic Thomson code is divided as follows: (1) six-dimensional phase space sampling of the relativistic electron beam, (2) classical electron dynamics for an electron in the phase space, and (3) calculation of the radiation across the electron beam phase space and three-dimensional laser focus. A linearly polarized laser pulse with a central wavelength $\lambda_0 = 800$ nm is used for the results presented in this paper. The polarization and propagation directions of the laser are parallel to the $x$- and $z$-axis, respectively.

### Phase Space Sampling of the Electron Beam

The electron beam, described by its average energy $E_0$, energy spread $\Delta E / E_0$, and divergence angle $\theta_0$, is sampled by a 6-dimensional phase space distribution. The phase space coordinate, $(x, y, z, p_x, p_y, p_z)$, of each electron in the beam is generated with a normal random number distribution of particles with respective standard deviations $\Delta p_x = \theta_e \times p_{z0}$, $\Delta p_y = \theta_e \times p_{z0}$, $\Delta x = \sigma_{xe}$, $\Delta y = \sigma_{ye}$, and $\Delta z = c \times \tau_e$, where $x, y, z$ are spatial coordinates, $p_x, p_y, p_z$ are the Cartesian components of the momentum, $\sigma_{x(y)e}$ horizontal (vertical) diameter of the electron beam, $\tau_e$ is the FWHM temporal duration of the electron beam, and $p_{z0}$ is the momentum corresponding to the average energy of the electron beam.

### Relativistic Electron Dynamics

Once the electron beam is described accurately by a sampled 6-dimensional phase space distribution, the dynamics of each electron in the laser field is calculated by solving the relativistic equations of motion given by,

$$\frac{d\vec{p}}{dt} = q\left(\mathbf{E}_{laser} + \frac{\vec{p}}{\gamma m_0} \times \mathbf{B}_{laser}\right)$$

$$\frac{d\vec{r}}{dt} = \frac{\vec{p}}{\gamma m_0},$$

where $m_0$ is the rest mass of an electron, $q$ the charge of an electron, **p** the particle momentum, $\mathbf{E}_{laser}$ and $\mathbf{B}_{laser}$ are the laser magnetic and electric field vectors, and $\gamma$ is the relativistic Lorentz factor.

The code we developed can calculate the laser field to a high degree of accuracy, i.e., non-paraxial field terms, relevant to a fast focus; terms up to order 7 are included. For simplification, the results presented in this paper are calculated with a plane wave laser field; allowing direct comparison with analytic solutions.

We use a 4$^{th}$-order Runge-Kutta ordinary differential equation solver with relative error-tolerance threshold of $10^{-6}$, a local error threshold of $10^{-12}$, and a time step typically of the order $10^{-4}$ femtosecond. Since the electron beams considered in this paper are relativistic and have low density ($n_e/\gamma^3 \ll 10^{16}$ cm$^{-3}$) space charge forces are neglected *(1, 17)*. In the absence of the laser field, the electron beam trajectory is assumed to be ballistic. Radiation damping is also not accounted for in these calculations, since the energy radiated per cycle by an electron is small compared to the energy of the electron.

### Radiation Calculation

Once the dynamics of each electron are obtained, the energy density radiated per unit frequency $\omega$ and solid angle $\Omega$ by a single electron moving in the intense laser field can be described by the classical formula *(18)*

$$\frac{d^2 I}{d\omega d\Omega} = 2|A(\omega)|^2,$$

where

$$A(\omega) = \left(\frac{e^2}{8\pi^2 c}\right)^{1/2} \int_{-\infty}^{\infty} e^{i\omega t}\left[\frac{\mathbf{n} \times [(\mathbf{n} - \boldsymbol{\beta}) \times \dot{\boldsymbol{\beta}}]}{(1 - \boldsymbol{\beta} \cdot \mathbf{n})^3}\right] dt,$$

where $e$ is the charge of an electron, $\omega$ angular frequency, $c$ speed of light in vacuum, **n** is a unit vector in the direction of observation, and **β** is the velocity of the electron normalized by the speed of light.

The total energy radiated per unit frequency $\omega$ and solid angle $\Omega$ by the electron beam, sampled with $N_e$ particles, is then obtained by summing over the entire phase space

$$\frac{d^2 I_{total}}{d\omega d\Omega} = \sum_{i=1}^{N_e} \frac{d^2 I_i}{d\omega d\Omega}.$$

## Benchmark Results and Discussion

To benchmark our code, we used the general analytic expressions of the radiated energy density, derived by integrating the Lienard-Wiechert potentials *(14, 19)*. Of particular interest is the limit of low strength laser fields, $a_0 < 1$, where the radiation is dominated by the first harmonic. For small observation angle $\theta$ and in the limit of non-relativistic scattering laser intensity ($a_0 \ll 1$), the energy radiated by a single electron per unit solid angle $\Omega$ and per unit frequency $\omega$ may be written as *(20, 21)*:

$$\frac{d^2 I}{d\omega d\Omega} = r_e m c \gamma^2 N_0^2 a_0^2 \left(\frac{\omega}{4\gamma^2 \omega_0}\right)^2 R(\omega, \omega_0),$$

where

$$R(\omega, \omega_0) = \left(\frac{\sin(\bar{k}L/2)}{\bar{k}L/2}\right)^2,$$

and $k$, $k_0$ are x-ray/laser wavenumbers, respectively, $\bar{k} = k(1+\gamma^2\theta^2)/(4\gamma^2) - k_0$, $L = N_0 \lambda_0$ is the interaction length, $N_0$ is the number of laser periods with which the electron interacts, $m$ is the rest mass of the electron, $c$ is the speed of light in vacuum, and $r_e = e^2/mc^2$ is the classical electron radius. For an electron beam with a finite energy spread and negligible divergence angle, the total scattered spectrum can be estimated analytically by summing over the electron beam energy distribution, with the spectrum from a single electron given by the above analytic formula. Typically, laser-wakefield accelerated electrons have large energy spread compared to the line-width of the scattered Thomson radiation from a single electron. In this case, the resonance function $R(\omega_x, \omega)$ can be approximated with a delta function, and the total radiated energy per unit frequency per unit solid angle may be obtained (20, 21) with,

$$\frac{d^2 I_T}{d\omega\, d\Omega} = \frac{1}{2} r_e m_e c N_0 a_0^2 \gamma^3 f(\gamma),$$

(2.1)

where $f(\gamma) = \frac{1}{N_e} \frac{dN}{d\gamma}$, $N_e$ is the total number of electrons in the beam.

The energy distribution of an electron beam sampled with $N_e = 10,000$ particles is shown in Fig. 1(a). The average energy of the beam is $\approx 100$ MeV, and the energy spread around the average energy is $27\%$. Initially the electron beam is located at $(x = 0, y = 0, z = z_e = 2c\tau)$, where c is the speed of light and $\tau$ is the temporal duration (FWHM) of the scattering laser pulse. An intense plane wave laser pulse, $I_0 = 9 \times 10^{16}$ W/cm$^2$ and $\lambda_0 = 800$ nm, located at $z = -z_e$ and travelling along $+z$-axis is backscattered by the electron beam, and the resulting backscattered energy density is shown in Figure 1 (b). The radiation calculated numerically with the developed code is in close agreement with the one obtained analytically with equation (2.1), see Figure 1(b). The peak radiated photon energy is 0.26 MeV, which might also be obtained with $E_{sc} = 4\gamma^2 \hbar \omega_0$.

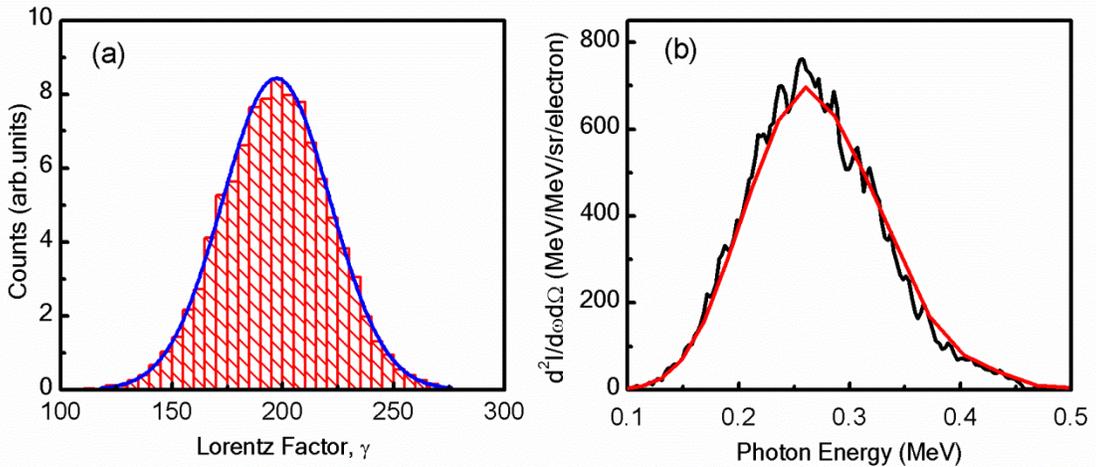

Fig. 1. Histogram plot of electron beam energy distribution (a), and the corresponding Thomson scattered radiated energy density (b) calculated numerically (black) and analytic estimate (red). The scattering laser intensity is $9 \times 10^{16}$ W/cm$^2$, $\lambda_0 = 800$ nm, and $\tau = 90$ fs. The electron beam is sampled with

$N_e = 10,000$. The energy spread of the electron beam is about $27\%$, and the corresponding Thomson bandwidth is $55.5\%$.

## Scattering from a Pulsed Laser and Spectral Broadening

Previous work investigated the effect of beam shapes on the Thomson scattered spectrum *(22-24)*. In particular, the pulsed nature of the laser pulse has been shown to introduce spectral substructures within the radiated harmonics. When an electron interacts with a pulsed laser pulse, it undergoes small oscillations during the rise of the pulse, where $a(t) \ll a_0$, and emits radiation at the relativistic Doppler shifted laser frequency $\omega_1$, see figure 2. When the laser field amplitude increases in magnitude, the ponderomotive force pushes the electron backward and it emits radiation which is frequency downshifted by a factor $1 + a_0^2/2$. When the field decreases back to zero, the electron radiates again at $\omega_1$ *(25)*. The spectral difference in the radiation from the increasing and decreasing parts of the laser pulse results in spectral interference of the radiated field, creating oscillations in the radiated spectrum *(24)*. In a recent study *(22)*, it is shown that the number of oscillations is proportional to the laser intensity and temporal duration, i.e., $N_\tau = 0.24 T_0(fs) a_0^2$. This non-linear oscillation of the spectrum should be accounted for when designing a narrow band gamma-ray source for precision applications such as nuclear resonance fluorescence.

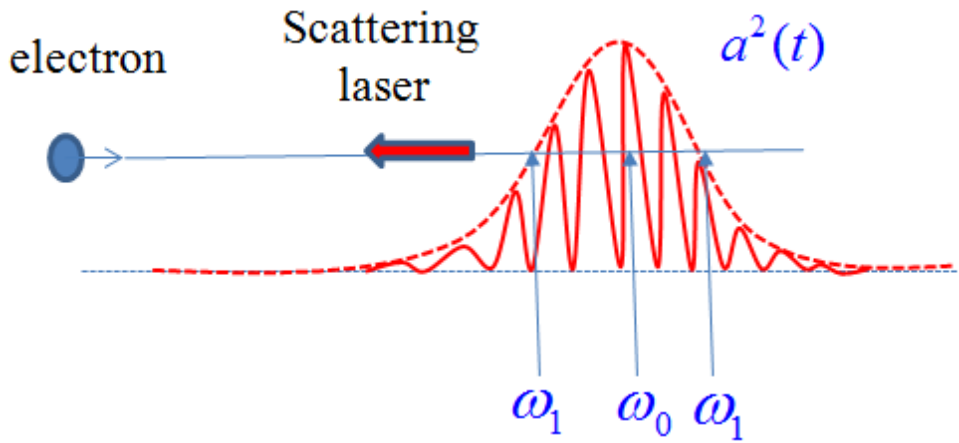

Fig. 2. Schematic diagram illustrating the time varying ponderomotive potential experienced by an electron as it collides with a counter-propagating laser pulse.

This non-linear oscillation in the spectrum might be minimized by using an appropriately chirped laser pulse. We propose a chirped laser pulse whose frequency changes with time as $\omega(t) = 2/3\omega_0(1+(a(t)/a_0)^2/2)$. This would ensure that radiated frequencies during the period of the laser pulse will be identical, and the spectral oscillations resulting from the spectral interference of the different radiated frequencies could be minimized. Since most of the radiation occurs near the peak of the laser intensity, the proposed chirp may be realized with a phase $\phi(t) = \omega_0 t - bt^3 + ct^5 + c.c$, where $b = \omega_0/(3\tau^2)$ is a third-order phase and $c = \omega_0/(30\tau^4)$ is a fifth order phase. For a 90 fs laser pulse, the amount of chirp required ($b^{-1} = 30903$ fs$^3$, $c^{-1} = 1.08 \times 10^8$ fs$^5$) can be produced with conventional stretcher/compressor and pulse-shaper (e.g. Dazzler) combinations. Figure 3 shows the radiated energy density ($d^2I/d\omega d\Omega$) when an 300-MeV electron is scattered off of a pulsed laser field with temporal duration 90 fs FWHM and peak intensity $1\times 10^{18}$ W/cm$^2$. The figure shows the spectral oscillations in the radiated spectrum is reduced in the chirped laser pulse case. For scattering from the transform limited laser pulse, the radiated fundamental harmonic is split into 12 substructures which are consistent with $N_\tau = 0.24 T_0(\text{fs})a_0^2$. For the chirped case, the radiated spectrum is dominated by the fundamental with only two less prominent substructures. Such a scheme might be used in the design of narrow band gamma-ray sources from the scattering relativistic electron beams with very small energy spreads, such as electron beams obtained with conventional accelerators.

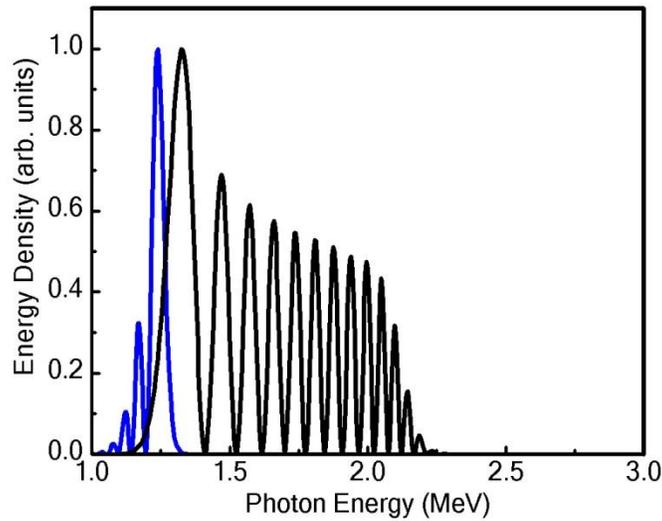

Fig. 3. Energy density from scattering of a 300-MeV electron by a 90-fs FWHM and $1 \times 10^{18}-W/cm^2$ peak intensity chirped laser pulse (blue line) and transform limited pulse (solid black line)

## Conclusion

In summary, based on the single particle trajectory tracking, a fully relativistic 3-dimensional non-linear Thomson scattering code has been developed and benchmarked against analytical expressions for scattering in a plane wave laser field. With a simple switch implemented in the code, the developed code is able to calculate the laser field to a high degree of accuracy. Furthermore, the code is used to investigate the beam shape effects on the scattered radiated energy. It was found that substructures in the emitted radiation spectrum (due to the pulsed nature of the laser pulse), which have a detrimental effect on the quality of the radiated spectrum, could be minimized by chirping the laser pulse appropriately.

## Acknowledgements

This material is based upon work supported by the U.S. Department of Energy: DE-FG02-05ER15663, Defense Threat Reduction Agency: HDTRA1-11-C-0001, Air Force Office for Scientific Research: FA 9550-08-1-0232, and FA9550-11-1-0157, Department of Homeland Security: 2007-DN-007-ER0007-02, and Defense Advanced Research Projects Agency: FA9550-09-1-0009. This work was completed utilizing the Holland Computing Center of the University of Nebraska. One of us (I.G) would also like to acknowledge very useful discussions with S.Y. Kalmykov and research group members of the Diocles Extreme-Light Laboratory at the University of Nebraska-Lincoln.